\documentclass[12pt,preprintnumbers,amsmath,amssymb,nofootinbib]{revtex4}
\usepackage[english]{babel}
\usepackage{multirow}
\usepackage{array}
\usepackage[letterpaper,top=2cm,bottom=2cm,left=3cm,right=3cm,marginparwidth=1.75cm]{geometry}

\usepackage{amsmath}
\usepackage{float}
\usepackage{graphicx}
\usepackage[colorlinks=true, allcolors=blue]{hyperref}
\newcommand {\nl} {\nonumber\\}

\begin{document}

\title{Composite nature of the $T_{cc}$ state}
\author{Xian-Wei Kang}
\author{Wen-Shuo Ding}
\affiliation{Key Laboratory of Beam Technology of Ministry of Education, School of Physics and Astronomy, Beijing Normal University, Beijing 100875, China\\
and Institute of Radiation Technology,
Beijing Academy of Science and Technology, Beijing 100875, China}

\begin{abstract}
In 2021, LHCb collaboration reported a very narrow state in the $D^0D^0\pi^+$ mass spectrum just below the $D^{*+}D^0$ mass threshold. We consider the influence of the Castillejo-Dalitz-Dyson (CDD) pole in the scattering amplitude to derive a general treatment for the two-body final state interaction near its threshold. The line shape (or the energy dependent event distribution) are then obtained, where the parameters can be fixed by fitting to the experimental data on the $D^0D^0\pi^+$ mass spectrum. Within our method the data are quite well reproduced. The pole structure in the complex energy plane indicates that the $T_{cc}$ state has a large portion of elementary degree of freedom (e.g., the compact tetraquark component) inside its hadron wave function. The compositeness as a measure of molecule component in its wave function is predicted to be $0.23_{-0.09}^{+0.40}$. Clearly, the non-molecular component takes a non-negligible or even dominant portion.
\end{abstract}
	
\maketitle
		
%\pagebreak[4]
		
%\tableofcontents
		
%\newpage

\section{Introduction}
In 2021, LHCb reported a narrow state named by $T_{cc}$ in $D^0D^0\pi^+$ mass spectrum just below the $D^{*+}D^0$ mass threshold. The pole in the second Reimann sheet of the $D^{*+}D^0$ scattering amplitude with respect to the $D^{*+}D^0$ threshold is found to be \cite{LHCb:2021vvq,LHCb:2021auc}
\begin{eqnarray}\label{eq:polepara}
\delta m_{\rm pole}&=&-360\pm40^{+4}_{-0}\,\,\text{keV},\nl
\Gamma_{\rm pole}&=&48\pm2^{+0}_{-14}\,\,\text{keV}.
\end{eqnarray}
Since the state is so close to the $D^{*+}D^0$ threshold, Breit-Wigner parameterization is not appropriate and we should focus this pole parameter as the mass and width values.

This observation arose great interest concerning its feature in the hadron world.  It was understood as the molecular state generated by $DD^*$ scattering \cite{Feijoo:2021ppq,Albaladejo:2021vln,Abolnikov:2024key,Du:2021zzh,Dai:2023cyo,Dai:2023kwv}, or the compact tetraquark \cite{Karliner:2017qjm,Kim:2022mpa,Weng:2021hje,Dong:2024upa,Ader:1981db,Heller:1986bt}, or the virtual state by a refined data analysis utilizing K-matrix with the Chew-Mandelstam formalism \cite{Dai:2021wxi}, or kinematic singularity \cite{Achasov:2022onn}. The lattice analysis assigns the observed $T_{cc}$ as the virtual state \cite{Whyte:2024ihh}. For a review one may refer to Ref.~\cite{Chen:2022asf}. In fact, a genuine state may involve several configurations. There are recent developments for constructing the compositeness to quantify the molecular component \cite{Guo:2015daa,Kinugawa:2023fbf,Baru:2003qq,Sekihara:2014kya,Aceti:2014ala}.

The Castillejo-Dalitz-Dyson (CDD) pole is proposed in the context of Low equation in 1956 \cite{Castillejo:1955ed}. They pointed out that infinite number of parameters (pole location and residue corresponding to CDD pole) can appear in the Low equation, and also those appearance of CDD pole signifies the amount of the molecular component. If the CDD pole sits very close to the threshold of the two scattering hadrons, elementary degree of freedom is dominant for the  generated resonances by the scattering, and otherwise, the two-body molecular component is dominant.

Combining the concept of the compositeness and the scattering amplitude with inclusion of the CDD pole has been an effective method to analyze the
inner structure of hadron state. The application along this line can be found in Refs.~\cite{Kang:2016jxw,Kang:2024gxg} for $X(3872)$; in Refs.~\cite{Kang:2016ezb,Zhang:2022hfa} for $Z_b(10610)$ and $Z_b(10650)$; in Ref.~\cite{Wang:2022vga} for $f_0(980)$; in Ref.~\cite{Guo:2016wpy} for $\Lambda_c^+(2595)$ and in Ref.~\cite{Gao:2018jhk} for other near-threshold heavy flavor resonances.

We next in Sec.~\ref{sec:fit} derive the expression for the partial-wave amplitude with the impact of the CDD pole, from which we derive the event distribution. The parameters are obtained by fit to the mass spectrum data. In Sec.~\ref{sec:result} we show our theoretical results with the analysis of the compositeness value. In Sec.~\ref{sec:conclusion} we render the summary and outlook.

\section{Scattering amplitude and event distribution}
\label{sec:fit}
Due to the extreme proximity between the state $T_{cc}$ and $D^{*+}D^0$ mass threshold, within one MeV, it is enough that to consider the $D^{*+}D^0$
scattering in non-relativistic form. The spin-parity quantum number of $T_{cc}$ is determined to be $1^+$, so the $D^{*+}D^0$ scattering mainly occurs in $S$ wave. For such case, $D^{*+}D^0$ scattering amplitude in terms of the energy of $DD^*$ can be written as \cite{Kang:2016jxw,Zhang:2022hfa,Guo:2016wpy}
\begin{equation}\label{eq:ts}
t(E)=\left[\frac{\lambda}{E-M_{\text{CDD}}}+\beta-\mathrm{i}k\right]^{-1},
\end{equation}
where $M_{\text{CDD}}$ and $\lambda$ are the position and residue, respectively, for the CDD pole; $m_{\text{th}}=m_{D^*}+m_{D^0}$ is the mass threshold; and $k=\sqrt{2\mu_{D^*D}(E-m_\text{th})}$ is the magnitude of three-momentum with $\mu_{D^*D}=m_{D^0}  m_{D^{\ast+}} /(m_{D^0} + m_{D^{\ast+}})$ denoting the reduced mass of $D^{*+}$ and $D^0$. The unitarity condition is $\text{Im}\, t^{-1}(E)=-k$.
In the second Riemann Sheet (RS), indicated by a superscript $\text{II}$, it becomes
\begin{equation}
t^{\text{II}}(E)=\left[\frac{\lambda}{E-M_{\text{CDD}}}+\beta+\mathrm{i}k\right]^{-1}.
\label{eqt2s}
\end{equation}
Note that there is a change of sign in front of $k$ comparing with $t(E)$ for the first (physical) RS in Eq.~\eqref{eq:ts}.
In this convention $k$ is always calculated such that $\text{Im}\,k>0$.
Comparing to the effective range expansion $t(E)=\left(-\frac{1}{a}+\frac{1}{2}r k^2 -\mathrm{i} k\right)^{-1}$, one can obtain the scattering length $a$ and effective range $r$ as
\begin{equation}
\begin{aligned}
\frac{1}{a}&=\frac{\lambda}{M_{\text{CDD}}-m_\text{th}}-\beta,\\
r&=-\frac{\lambda}{\mu_{D^*D} (M_{\text{CDD}}-m_\text{th})^2}.
\end{aligned}
\label{eqar}
\end{equation}
Clearly, when $M_{\text{CDD}}\approx m_{\text{th}}$, i.e., the CDD pole position is extremely close to the two-meson threshold,  $r$  goes to infinity. In this circumstance, the effective range expansion approach fails due to the too limited convergence radius. We do encounter this situation of $r\to \infty$ in Refs.~\cite{Kang:2016jxw} and \cite{Zhang:2022hfa}. Or more precisely, the data quality is not capable of pinning down the location of CDD pole.

As a matter of the fact, CDD pole is zero of the scattering amplitude $t(E)$, which strongly distorts the energy dependence of $t(E)$ near $E\approx M_{\text{CDD}}$. The production process is mediated by the following $d(E)$ (but not $t(E)$) by removing the extra $E-M_{\text{CDD}}$ factor in $t(E)$ \cite{Guo:2016wpy}:
\begin{equation}\label{eq:dEgeneral}
d (E) = \left(1 + \frac{E - M_{\text{CDD}}}{\lambda} (\beta - \mathrm{i} k)\right)^{-1}
\end{equation}
We use the function $d(E)$ to treat the final state interaction. The factor $|d(E)|^2$ constituents the parametrization for the signal. Near the resonance region, one has the form
\begin{equation}\label{eq:dEpole}
d(E)\simeq\frac{\gamma_E}{E-E_P},
\end{equation}
with $E_P=M_P-\mathrm{i}\,\Gamma_P/2$ and $\gamma_E$ denoting the pole and its residue in the complex $E$-plane. The residue can be calculated by an integration along a closed contour around the pole:
\begin{equation}
\gamma_E=\frac{1}{2\pi i}\oint d(E)\mathrm{d}E.
\end{equation}
Then the mass distribution can be written as
\begin{equation}\label{eq:dMdE}
\frac{\mathrm{d} N}{\mathrm{d} E}=\frac{\Gamma_P|d(E)|^2}{2\pi|\gamma_E|^2},
\end{equation}
with $\Gamma_P$ denoting the pole width. We take the normalization condition such that for a narrow resonance case, the following integration
\begin{equation}
\mathcal{N}=\int_{-\infty}^{+\infty} \mathrm{d} E \frac{\mathrm{d} N}{\mathrm{d}E}
\end{equation}
is close to 1. In such scenario, the following $Y_{\text{const}}$ can be understood as the yield for this decay process. Similar choice is also taken in Refs.~\cite{Kang:2016jxw,Braaten:2009jke}. When $E_P$ corresponds
to a virtual state or other situations for which the final-state interaction function $d(E)$ has a shape that
strongly departs from a non-relativistic Breit-Wigner one, $\mathcal{N}$ could be far away from 1. Our description Eq.~\eqref{eq:dEgeneral} can display various line shapes beyond one shown in Eq.~\eqref{eq:dEpole} and provides a more general treatment for the final state interaction. So we claim Eq.~\eqref{eq:dEgeneral} is applicable in a wider range.

We now consider the experimental mass resolution and the corresponding background.  Following the experiment \cite{LHCb:2021auc}, the energy resolution function is described by a sum of two Gaussian functions
\begin{equation}
 R(E^{'},E) = \frac{0.778}{\sqrt{2 \pi} \sigma_1} \exp\left[\frac{(E^{'} - E)^2}{-2 \sigma_1^2}\right] + \frac{0.222}{\sqrt{2 \pi} \sigma_2} \exp\left[\frac{(E^{'} - E)^2}{-2 \sigma_2^2}\right],
\end{equation}
with $\sigma_1=276.15$ keV and $\sigma_2=666.35$ keV. The background contribution is parameterized by
\begin{equation}
B(E)= P_2\times\varPhi_{D^{\ast+}D^0},
\end{equation}
with $P_2=a E^2 +b E+ c$ denoting a positive second-order polynomial, and
\begin{equation}
\varPhi_{D^{\ast+}D^0}=\frac{\sqrt{(E^2 - (m_{D^0} + m_{D^{\ast+}})^2)(E^2 - (m_{D^0} - m_{D^{\ast+}})^2)}}{2E}
\end{equation}
the two-body phase space factor. Finally, we obtain the energy-dependent event number distribution in an energy bin of width $\Delta = 500$ keV centered at $E_i$:
\begin{equation}
N(E_i)= \int^{E_i+\Delta/2}_{E_i-\Delta/2} \mathrm{d}E'\int_{-\infty}^{+\infty} \mathrm{d}E \left[Y_{\text{const}} \left|d(E)\right|^2 +B(E)\right] R(E',E).
\end{equation}
We have 7 parameters in total: $\lambda, M_{\text{CDD}}, \beta$ in $d(E)$ function, $a,b,c$ in the background, and $Y_{\text{const}}$ as the overall normalization constant. These parameters could be fixed by fit to the data. In the fit, we should include the finite $D^{*+}$ width, $\Gamma_{D^{*+}}=83.4$ keV \cite{ParticleDataGroup:2024cfk}, and as a result,
 \begin{equation}
k = \sqrt{2 \mu_{D^*D} (E-m_{\text{th}} + \mathrm{i} \Gamma_{D^{*+}}/2)}.
\end{equation}
The $T_{cc}$ state is so close to threshold (recalling the value of $\delta m_{\text{pole}}$ in Eq.~\eqref{eq:polepara}) such that the inclusion of 83.4 keV is necessary, as has also been verified by the numerical calculation.

\begin{figure}[htbp]
    \includegraphics[width=\textwidth]{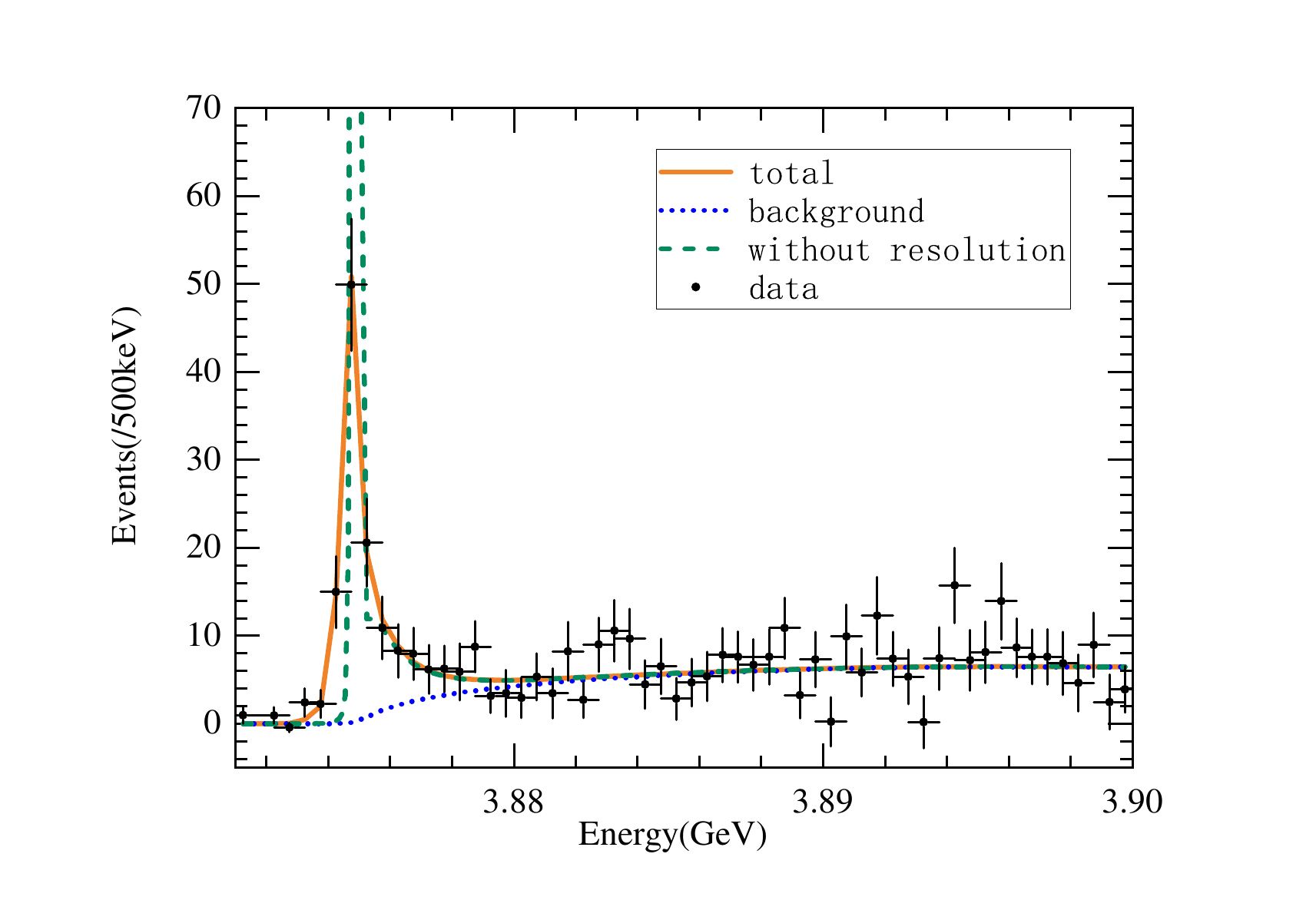}
    \caption{Mass spectrum for the $D^0D^0\pi^+$ decay channel. The data are from the LHCb collaboration \cite{LHCb:2021auc}. The solid line represents our total result, with dotted line showing the background and dashed line corresponding to without the Gaussian resolution.}\label{fig:fig7n}
\end{figure}

We use the optimization algorithm of MINUIT package to perform the minimization of $\chi^2$ \cite{MINUIT}. The parameter values are shown in Table
\ref{tab:fit}. In the calculation, all the mass, energy and momentum are expressed in units of MeV. Then the units for the parameters are also given in the table. The fit result is shown in Figure \ref{fig:fig7n}, where ``Energy'' corresponds to the invariant mass of $D^0D^0\pi^+$. In fact, near the pole (signal) region, the invariant mass $M_{D^{*+}D^0}\approx M_P\approx M_{D^0D^0\pi^+}$. From the following discussions, we claim the $T_{cc}$ state can be understood as the $D^{*+}D^0$ bound state. For a bound state case, the aforementioned $\mathcal{N}\approx 1$ is clear since Lorentzian distribution reduces to a delta function. $Y_{\text{const}}$ can then be understood as the yield corresponding to the product of production cross section of $T_{cc}$ and branching fractions of $T_{cc}\to D^{*+}D^0$ and $D^{*+}\to D^0\pi^+$ (note that cross section has dimension of [MeV]$^{-2}$).

\begin{table}[htbp]
  \centering
    \begin{tabular*}{\columnwidth}{@{\extracolsep{\fill}}lll}
    \hline\hline
	$Y_{\text{const}} =(10.8 \pm 8.0)$ MeV$^{-2}$ &  & \\
	$\lambda=(83.6 \pm 63.8)$ MeV$^2$   &$\beta=(70.5 \pm 37.6)$ MeV & $M_{\text{CDD}}-m_{\text{th}}=(0.47 \pm 0.38)$ MeV  \\
	$a=-(81.4 \pm 2.8)$ MeV$^{-2}$  &$b=-(99.2 \pm 10.8)$ MeV$^{-3}$  &$c=(1653.5 \pm 41.9)$ MeV$^{-4}$\\
	\hline\hline
\end{tabular*}
 \caption{Our best values of the 7 fitted parameters, with $\chi^2$ per degree of freedom being 0.93. }
     \label{tab:fit}
 \end{table}

\section{Scattering parameter, pole and compositeness}
\label{sec:result}
Using the central values of the fitted parameters, one obtains the scattering length $a=1.8$ fm and the effective range $r=-77.2$ fm. The value of $r$ is so large that the effective range expansion method is not an appropriate tool in this case. Within uncertainties we have another set of parameter value, $ \lambda= 19.8,\,\beta= 108.0,\, M_{\text{CDD}}= 0.47+m_{\text{th}}$, which leads to $a=-3.0$ fm, $r=-18.3$ fm. The values of $r$ are not the typical magnitude of a few fermi. So we will concentrate on the outcome of our CDD analysis.

The definition of compositeness for a bound state is well defined by the Weinberg formula \cite{Weinberg:1962hj,Weinberg:1965zz}. However, for resonance case, the compositeness calculated in that way will become a complex-valued number. As mentioned before, there are several developments for defining the compositeness.  We will follow the one developed in Ref.~\cite{Guo:2015daa} for a resonance:
\begin{equation}
\label{eq:Xres}
X=|\gamma_s|^2\left| \frac{d G(s)}{d s}\right|_{s=E_P^2},
\end{equation}
for the case of $\sqrt{\text{Re} E_P^2}$ larger than the lightest threshold for a coupled-channel scattering.
$-\gamma_s^2$ is the residue of $t(s)$ at the resonance pole position, $E_P^2$, of the complex $s$ plane in the second RS:
\begin{equation}
t^{\text{II}}(s)\rightarrow \frac{-\gamma_s^2}{s-E_P^2}.
\end{equation}
$\gamma_s^2$ is related to $\gamma_E$ by \cite{Zhang:2022hfa,Kang:2016jxw}
\begin{equation}
\label{eq:gammas}
\gamma_s^2=-16\pi\gamma_E E_P^2\times\frac{E_P-M_\text{CDD}}{\lambda}.
\end{equation}
$G(s)$ is the two-point Green function, and the infinity can be removed by cutoff regularization or the dimensional regularization \cite{Kang:2016zmv}. Its form in the dimensional regularization scheme at the regularization scale $\mu$ can be written as \cite{Guo:2016wpy}
\begin{equation}
G(s)=\alpha(\mu^2)+\frac{1}{(4\pi)^2}\left(\log\frac{m_2^2}{\mu^2}-\varkappa_+\log\frac{\varkappa_+-1}{\varkappa_+}
-\varkappa_-\log\frac{\varkappa_--1}{\varkappa_-} \right),
\label{eqgs}
\end{equation}
with
\begin{equation}
\begin{split}
\varkappa_{\pm}&=\frac{s+m_1^2-m_2^2}{2s}\pm\frac{p}{\sqrt{s}}\,,\\
p&=\frac{\sqrt{(s-(m_1-m_2)^2)(s-(m_1+m_2)^2)}}{2\sqrt{s}}\,,
\end{split}
\end{equation}
It is equavalent to the one used in Ref.~\cite{Oller:2006jw}. The $G(s)$ defined in Eq.~\eqref{eqgs} corresponds to the physical value on the real axis, or $G^\text{I}(s+\mathrm{i}\epsilon)$ in the physical sheet.

Resonance appears as a pole in the second RS. We thus need to make an analytical extrapolation of $G(s)$ to the second RS. The result is given by
\begin{equation}
\label{eq:Gcon}
G^{\text{II}}(s+\mathrm{i}\epsilon)=G^{\text{I}}(s+\mathrm{i}\epsilon)+\frac{\mathrm{i}}{4\pi\sqrt{s}}\frac{\sqrt{(s-(m_1-m_2)^2)(s-(m_1+m_2)^2)}}{2\sqrt{s}}.
\end{equation}
The left and right side of Eq.~\eqref{eq:Gcon} implies the condition of $\text{Im}\,s > 0$. For $\text{Im}\,s < 0$ case, using the Schwarz reflection principle $G(s^*)=[G(s)]^*$, one has
\begin{equation}
\label{eq:Gcon2}
G^{\text{II}}(s-\mathrm{i}\epsilon)=G^{\text{I}}(s-\mathrm{i}\epsilon)
-\frac{\mathrm{i}}{4\pi\sqrt{s}}\frac{\sqrt{(s-(m_1-m_2)^2)(s-(m_1+m_2)^2)}}{2\sqrt{s}}.
\end{equation}
It is $G^{\text{II}}$ that should be used in Eq.~\eqref{eq:Xres} for a resonance. For a bound state ($E_B$) case, one simply has
\begin{equation}
\label{eq:Xbound}
X=-\gamma_s^2\left.\frac{d G^{\text{I}}(s)}{d s}\right|_{s=E_B^2},
\end{equation}
and of course $\gamma_s$ is calculated in the physical sheet. The situation of $X=1$ corresponds to a pure bound state or molecular. The elementariness $Z=1-X$ measures the weight of all other components in the hadron wave function. In fact, the non-relativistic form of $G(s)$ is
$\frac{1}{8\pi m_{\text{th}}}(\beta-\mathrm{i} k)$, and then to be consistent in the non-relativistic frame, one has
\begin{eqnarray}\label{eq:Xres-non}
X=\frac{1}{8\pi m_{\text{th}}}\left|\frac{\mu \gamma_s^2}{2E_Pk_P}\right|,
\end{eqnarray}
with the pole momentum $k_P=\sqrt{2\mu(E_P-m_{\text{th}})}$. In a real numerical calculation, such difference is negligible, for e.g., $X=0.232$ deducing from Eq.~\eqref{eq:Xres} versus $X=0.232$ from Eq.~\eqref{eq:Xres-non}.

We search for the pole in the complex energy plane. We first consider the well-defined elastic scattering case with $\Gamma_{D^{*+}}\to 0$. As a result, for the central value of the fitted parameter we find a pole 3874.72 MeV in the physical sheet, as a bound state; and a pole $3874.48 - \mathrm{i}\, 1.74$ MeV in the unphysical Riemann sheet, as a resonance. By slowly varying e.g., the values of $M_{\text{CDD}}$, both poles move gradually. For the bound state pole, we have the residue $\gamma_s^2=5.04$ GeV$^2$, and the corresponding compositeness $X=0.23$.
Considering the uncertainties in Table.~\ref{tab:fit} within one sigma region, we have $\gamma_s^2=5.04^{+2.16}_{-1.60}$ GeV$^2$ and $X=0.23_{-0.09}^{+0.40}$. As for the uncertainty, it is always hard to be quantified well. In the current calculation, we take the uncertainty from MIGRAD method in MINUIT subroutine. On the other hand, the propagation of the error for the parameter to the final $X$ value is highly nonlinear, since we need to search for the corresponding pole and calculate its residue for each parameter set. We discretize the parameter values into dozen of sets and then get dozen of poles, residues, and $X$. Then the central value, upper value, lower value are extracted. The parameter set $\lambda=19.8,\beta=33.0,M_{\mathrm{CDD}}=0.47$ in Table I gives $X=0.63$, which leads to the largest uncertainty. The LHCb collaboration uses the Weinberg compositeness formula involving the values of the scattering length and effective range, to investigate the elementariness $Z$ value. The result is $Z< 0.52 \, (0.58)$ at 90 (95)\% confidence level. So it implies the compositeness $X>0.46$ at 90\% confidence level, which overlap with our value of $X$ in some range. However, much work towards the detailed study of $X$ still needs to be done.

The finite but small $D^{*+}$ width moves the aforementioned resonance pole to $3874.5 -\mathrm{i} 1.73$ MeV and the bound state pole to $3874.72 - \mathrm{i}\,0.0098$ MeV. It is interesting to note that our such finding of a bound state pole agrees well with Ref.~\cite{Lin:2022wmj}, where they find a bound state with width of 80 keV by using the complex scaling method. Using the parameter values $\lambda=19.8,\,\beta=108.0,\, M_{\text{CDD}}=0.47+m_{\text{th}}$ we can achieve a bound state pole at $3875.4 - \mathrm{i}\, 0.039$ with the width of 78 keV for the unstable bound state. If the condition $\sqrt{\text{Re}\,E_P^2}\geq m_{\text{th}}$ is fulfilled for a resonance pole, the compositeness formula, Eq.~\eqref{eq:Xres}, could be applied and the compositeness $X$ ranges from 0.07 to 0.57 accordingly. In fact, given the fact that both poles appear in the physical and unphysical sheet, $T_{cc}$ should rather be an elementary state, e.g., the compact tetraquark component takes a large or even dominant portion, according to the Morgan criterion \cite{Morgan:1992ge}.

For a fit, one usually worries the local minimum. We illustrate this briefly here. In Ref.~\cite{Zhang:2022hfa}, we have imposed a pole in the second sheet with pole parameters given by the experimental determination. In this way, $\lambda$ and $\beta$ are expressed as functions of $M_\text{CDD}$, and then two parameters will be reduced. We have tried the similar exercise in our present study. The fit result is acceptable, but not as good as the one above. One of the reason could be that the width of $T_{cc}$ is much narrower or the peak is more prominent than the ones for $Z_b$ states. We will not show its parameter value and figure. However, the conclusion that both poles in the first and second sheet are found does not change. In the second sheet, the pole is at $-0.36 \, \text{MeV}+m_{\text{th}}-\mathrm{i}\,0.024 \,\text{MeV}$ (cf.~Eq.~\eqref{eq:polepara}), as required. In this case we can not provide its compositeness value since it locates below the threshold.  For the bound state case, a small compositeness number is found, which implies the compact tetraquark component in its wave function can not be overlooked. That is, this fit solution does not influence our main conclusion at all. Certainly, we prefer the solution that is presented in the Table I and Figure 1.

\section{Conclusion and outlook}
\label{sec:conclusion}
The mystery of $T_{cc}$ state reported by LHCb has not been unveiled, where various interpretations are proposed. We utilize an amplitude including the CDD pole to incorporate the nontrivial and non-perturbative final state interaction of $D^{*+}D^0$. This method provides much more information than the effective range expansion method. The appearing parameters are $\lambda,\,\beta$ and $M_{\text{CDD}}$, which will be fixed by fitting to the data. In the fit the experimental energy bin width and mass resolution function are considered. With the known parameters, we search for the pole in the complex energy plane. Both poles are found in the physical and unphysical Riemann sheet. This indicates that $T_{cc}$ can be interpreted as an elementary state, e.g., the compact tetraquark component takes a large portion. The coupled-channel study following Ref.~\cite{Kang:2016jxw} may render more useful information and this work is ongoing.

\section*{Acknowledgments}
We are indebted to J.~A.~Oller for helpful discussion. We also acknowledge the discussion with Han-Qing Zheng on the Morgan pole counting rule.
This work is supported by the National Natural Science Foundation of China under Project No.~12275023.

%\bibliographystyle{alpha}
%\bibliography{sample}

\end{document}